\documentclass[journal]{IEEEtran}

\usepackage{adjustbox}
\usepackage{color}
\usepackage{epsf}
\usepackage{times}
\usepackage{epsfig}
\usepackage{epstopdf}
\usepackage{graphicx}
\usepackage{algorithm}
\usepackage{algorithmic}
\usepackage{amsmath}
\usepackage{amssymb}
\usepackage{amsxtra}
\usepackage{enumerate}
\usepackage{multirow}
\usepackage{float}
\usepackage{mathtools}
\usepackage{verbatim}
\usepackage{subcaption}
\usepackage{multirow}
\usepackage{comment}
\usepackage[normalem]{ulem}
\usepackage{tabu}
\usepackage[dvipsnames]{xcolor}
\usepackage{multicol}
\usepackage{cite}
\usepackage{caption}

\newcommand\blfootnote[1]{%
  \begingroup
  \renewcommand\thefootnote{}\footnote{#1}%
  \addtocounter{footnote}{-1}%
  \endgroup
}

\usepackage{balance}

\usepackage{titlesec}
\titlespacing*{\section} {0pt}{3ex}{2ex}
\titlespacing*{\subsection} {0pt}{2ex}{1ex}
\titlespacing*{\subsubsection} {0pt}{2ex}{1ex}
\usepackage[skip=5pt, indent=20pt]{parskip}
\patchcmd{\thebibliography}
  {\settowidth}
  {\setlength{\parsep}{1pt}\setlength{\itemsep}{1pt plus 0.1pt}\settowidth}
  {}{}

\begin{document}

\title{Enhanced In-Flight Connectivity for Urban Air Mobility via LEO Satellite Networks}

\begin{comment}
\author{
\IEEEauthorblockN{Karnika Biswas,  Hakim Ghazzai, \textit{Senior Member, IEEE}, Abdullah Khanfor, \textit{Member, IEEE}, and Lokman Sboui, \textit{Senior Member, IEEE}}\\
{\thanks {\hrule
\vspace{0.1cm} 
Karnika Biswas, Hakim Ghazzai, are with the Computer, Electrical, Mathematical Science and Engineering Division (CEMSE), King Abdullah University of Science and Technology (KAUST), Thuwal, Saudi Arabia. (E\textendash mails: \{karnika.biswas,hakim.ghazzai\}@kaust.edu.sa).\newline
Abdullah Khanfor is with the Computer Science Department, College of Computer Science \& Information Systems, Najran University, Najran, Saudi Arabia.
(E\textendash mail: aikhanfor@nu.edu.sa)\newline
Lokman Sboui is with the Systems Engineering Department, \'Ecole de Technologie Sup\'erieure (\'ETS), University of Qu\'ebec, Montr\'eal, Canada.\newline
}}}
\end{comment}

\author{\IEEEauthorblockN{Karnika Biswas$^{1}$, Hakim Ghazzai$^{1}$, Abdullah Khanfor$^{2}$ and Lokman Sboui$^{3}$}\\\vspace{0.15cm}
\IEEEauthorblockA{
\small 
$^{1}$CEMSE Division, King Abdullah University of Science and Technology (KAUST), Thuwal, Saudi Arabia\\\vspace{0.1cm}
$^{2}$Computer Science Department, College of Computer Science \& Information Systems, Najran University, Najran, Saudi Arabia\\\vspace{-0.1cm}
$^{3}$Systems Engineering Department, \'Ecole de Technologie Sup\'erieure (\'ETS), University of Qu\'ebec, Montr\'eal, Canada\\
%Email: karnikabiswaschandra@gmail.com, hakim.ghazzai@kaust.edu.sa, aikhanfor@nu.edu.sa, lokman.sboui@etsmtl.ca
}}
\maketitle
\thispagestyle{empty}

\begin{abstract}
Urban Air Mobility (UAM) is the envisioned future of inter-city aerial transportation. This paper presents a novel, in-flight connectivity link allocation method for UAM, which dynamically switches between terrestrial cellular and Low Earth Orbit (LEO) satellite networks based on real-time conditions. Our approach prefers cellular networks for cost efficiency, switching to LEO satellites under poor cellular conditions to ensure continuous UAM connectivity. By integrating real-time metrics like signal strength, network congestion, and flight trajectory into the selection process, our algorithm effectively balances cost, minimum data rate requirements, and continuity of communication. Numerical results validate minimization of data-loss while ensuring an optimal selection from the set of available above-threshold data rates at every time sample. Furthermore, insights derived from our study emphasize the importance of hybrid connectivity solutions in ensuring seamless, uninterrupted communication for future urban aerial vehicles.
\end{abstract}

\blfootnote{\hrule
\vspace{0.1cm} This paper is accepted for publication in The 30th IEEE International Conference on Telecommunications (ICT2024), Amman, Jordan, Jun. 2024. \newline \textcopyright~2024 IEEE. Personal use of this material is permitted. Permission from IEEE must be obtained for all other uses, in any current or future media, including reprinting/republishing this material for advertising or promotional purposes, creating new collective works, for resale or redistribution to servers or lists, or reuse of any copyrighted component of this work in other works.}%

\section{Introduction}
\label{intro}

Urban Air Mobility (UAM) has gained substantial attention in recent years as a promising solution for mitigating urban congestion and revolutionizing inter-city transportation. At its core, UAM promises a zero-emission footprint and optimized utilization of resources, making it an environmentally sustainable and efficient means of transport~\cite{uamoverview}. As urban ecosystems rapidly evolve, there is a growing need for aircraft to be environmentally friendly but also safe, noiseless, and cost-competitive carriers. The modern aircraft envisioned for UAM operations embodies these characteristics. However, the promise of UAM is accompanied by challenges that need to be addressed to be effectively integrated into urban landscapes and accepted by the public~\cite{uamchallenge}. Foremost among these is the on-demand availability and operability of systems such as infrastructure and services, and reliable passenger data-connectivity, constituting the vital in-flight and ground support environment for UAM~\cite{uamchallenges2}.

%\textbf{paragraphs 1 and 2 should be combined into a single and short paragraph highlighting the emergence of UAM and the associated challenges }

From a technological perspective, connectivity is a crucial feature for UAM traffic management, monitoring, and passenger experience~\cite{10145033}. The critical nature of data exchange in a UAM system requires uninterrupted and reliable connectivity for efficient routing and encompasses various layers of communication. From UAM vehicle to control stations, seamless data exchange is essential, in particular, the Machine-to-Machine (M2M) communication, enabling the envisioned swarm operations using Wireless Mesh Networks (WMN)~\cite{survey_droneswarm}. 
Furthermore, navigation is UAM's operational integrity, alongside ensuring the safety of passengers and the urban populace. Thus, it must involve mechanisms to handle urgent landing scenarios and establish tactical deconfliction strategies, thereby negating the risks of mid-air collisions or accidents~\cite{uamnavigation}. However, the absence of cellular network coverage in sparsely populated regions and at the UAM's altitude of travel poses the greatest challenge in connectivity.

%\textbf{paragraphs 3 and 4 should be combined into a single paragraph discussing the importance of connectivity (for navigation, monitoring, and passenger leisure) and its challenges (absence of connectivity at high altitude)  }

Therefore, to provide the in-flight essential communication services for fault detection and broadcast, system malfunction, routing, resource allocation, and data security in non-covered areas and at high altitudes, UAM needs an alternative communication network. Low Earth Orbit (LEO) satellites are unanimously accepted as a good option. However, standalone usage of the LEO satellite network could not be more practical due to limited throughput, high cost, low data rate, and intermittent satellite visibility. Therefore, the authors of~\cite{UAVLEO} have proposed a combination of cellular and satellite networks to maximize network coverage.

%\textbf{A new paragraph should present LEO sats and their importance and how they can be used with UAM (provide connectivity in non-covered areas and at high-altitude, highlight that leo sat throughput is limited  and highlight the issue of mobility of LEO)}

%\textbf{Another new paragraph should discuss at least 3 related work, you can highlight that there is no work on UAM/LEO (if you dont find) and talk about some strategies using LEO sats for UAVs or HAPs.}\par
UAM, being a relatively new research field, the combination of using LEO satellites with cellular networks, specifically UAM, is not widely available. However, extensive research has already been reported for UAVs, such as devising a dynamic learning-based resource allocation after estimating energy consumption~\cite{resource_alloc} and increasing data-offloading throughput using a terrestrial satellite terminal with time division multiplexing~\cite{ultradenseLEO}. In~\cite{XiA}, a delay-aware routing algorithm through LEO and 5G networks has been proposed to facilitate time-sensitive information exchange with UAVs; these concepts can be also applicable to UAM vehicles.

In this paper, we investigate an in-flight connectivity framework for UAM using a hybrid communication system where an eVTOL is able to communicate with either cellular networks or LEO satellites. After defining the system model, we develop an optimization problem to determine the best link for communication among the available satellites and Base Stations (BSs). To simulate the  problem, we develop a scenario using the Satellite Communication Toolbox of Matlab and analyze the communication performance along the eVTOL trip in terms of achievable rate and data loss. We show that LEO satellite can play an important role in maintaining connectivity and reduce data loss during high-altitude flights or in remote area where ground connectivity is unavailable.   

%\textbf{I will comment/modify the contribution paragraph at the end}
\section {System model}
%\textbf{we need to give an index to the time slots, it seems in the next section, it is called $k$ in that case we should add the index $k$ to everything here, or simply at the end of the section II, mention that index k for time slots will be added to all time-depedent variables, e.g., $r_i^{cell}(k)$. All the time notations should follow like $\Delta(k)$ not $\Delta_k$, please also explain what $\delta t$ the duration fot eh time slot}
In the following paragraphs, we describe the system model assuming that a UAM trip is divided into $k$ time-slots. A time-slot is equivalent to one sampling interval, during which the corresponding link data-rate is assumed to remain unchanged. Data-rates for communication links are evaluated at each time slot, thereby all time-dependent variables and time notations follow the $k$ index, where $k \in \{0,1,2,..., \check{K}\}$. The duration of a time slot between two consecutive time-samples is denoted by $\delta t$.  where $\delta t$ is the time-slot duration. 

\subsection{Network Architecture}
We consider an eVTOL flying between two distant cities following a pre-determined trajectory, similar to commercial flights, but at a lower altitude and average velocity. Typically, an eVTOL reaches an altitude of 600-1800 meters above sea level and travels at 110-300 km/hr. According to its trajectory and speed, the trip lasts for a pre-determined duration. We assume that the eVTOL will communicate with external entities, e.g., air traffic control, other distant eVTOLs, and/or mobile operators, through cellular networks or LEO satellite links. To this end, we consider a set of cellular BSs distributed along the eVTOL route. The geodetic locations of these BSs are denoted by $(x_i,y_i)$ where $i=1,\dots,N_{cell}$ with $N_{cell}$ the number of BSs to which the eVTOL could connect during its trip. The BSs are all assumed to be at the same height $z_i$. The UAM is assumed to be equipped with one omni-directional antenna. \par
\begin{comment}
Figure~\ref{Fig1} illustrates the distribution of BSs along the eVTOL route, which we consider a typical cellular network where BS antennas are tilted towards the ground. 
\begin{figure}[!h]
\centering
{\includegraphics[width=9cm]{Fig1.png}} 
\caption{eVTOL trajectory and the spatial distribution of cellular BSs. }
\label{Fig1}
\end{figure}
\end{comment}

\begin{figure}[t]
\centering
{\includegraphics[width=8cm]{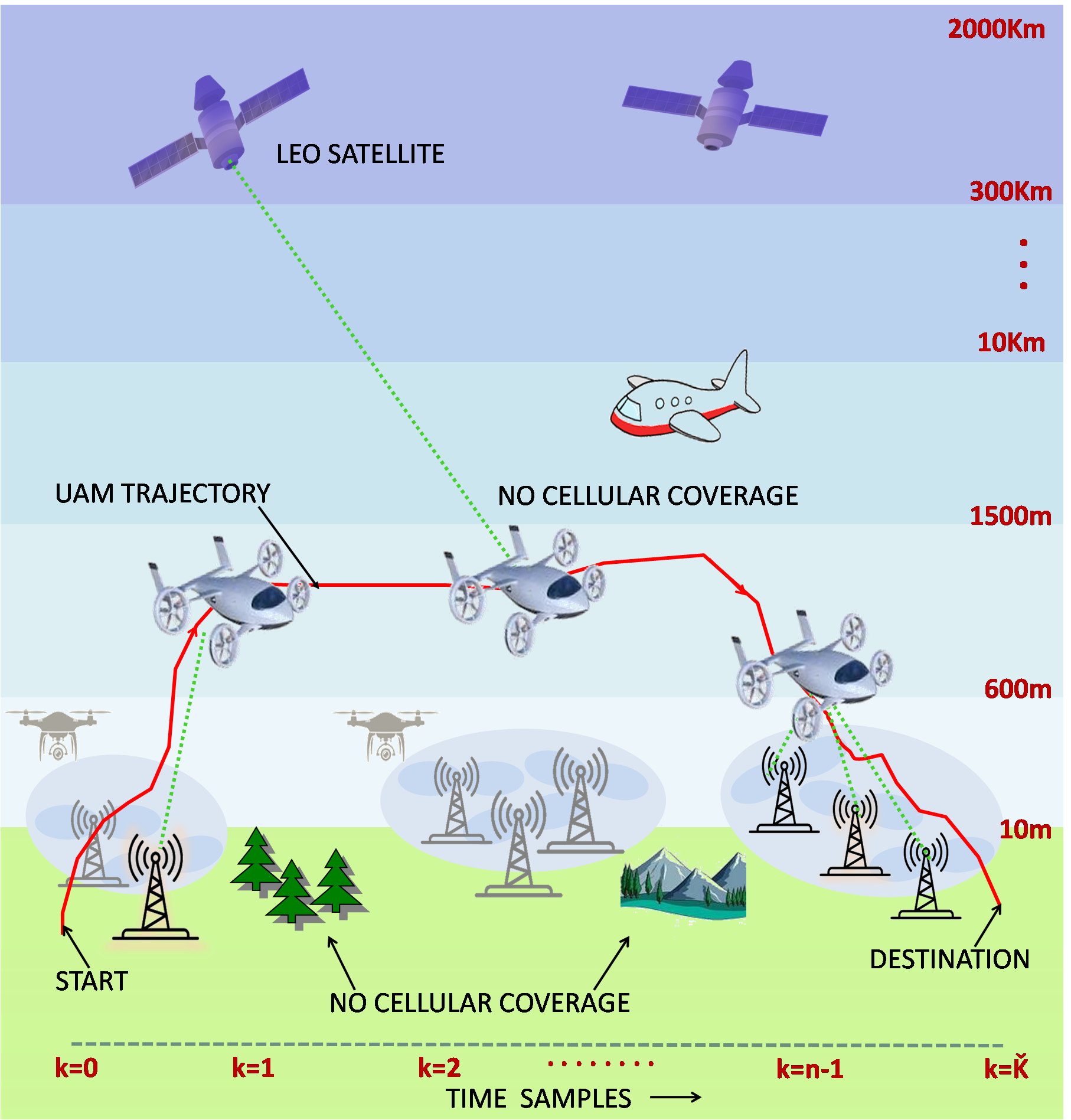}} 
%{\includegraphics[width=15cm]{Figure1.png}} 
\caption{UAM connectivity to cellular and satellite networks in a trip. }
\label{Fig2}
\end{figure}

On the other hand, we consider a constellation of LEO satellites with known motion patterns, each orbiting the Earth at a fixed altitude between 340-614 km. The satellite poses are assumed to be available at each time sample in a Geocentric Celestial Reference Frame (GCRF) format, from which the geographic coordinates  $(u_j,v_j, w_j)$ where $j=1,\dots,N_{sat}$ with $N_{sat}$ being the maximum constellation size are computed. A uniform Latitude-Longitude-Altitude (LLA) representation of positions facilitates the computation of the eVTOL-satellite distance and eVTOL-BS distance. Figure~\ref{Fig2} schematically shows how UAM-satellite and UAM-cellular links are established at different waypoints of a trip. 
%Cellular communication is the preferred mode at lower altitudes during take-off and landing, whereas, during high-altitude stable flight, LEO satellite communication is the desired connectivity link.

\subsection{Channel Models}
\subsubsection{UAM-Cellular Channel Model}
Cellular-eVTOL communication is mainly considered when the eVTOL is under the coverage of the BSs. i.e., at a certain low altitude under which the Signal-to-Noise Ratio (SNR) is sufficient for wireless communication. The eVTOL may have multiple cellular links available at a given time instant. A cellular channel is assumed to follow a known multi-path signal propagation model. Therefore, the channel data rate can be computed at each sampling instant using the knowledge of the separation distance between the eVTOL and the BS, the carrier frequency, and the average propagation losses. We also assume that the Channel State Information (CSI) and the computed data rate remain independent and constant during the transmission~\cite{channelmodel}.

%\footnote{ Data update time is assumed higher than channel coherence time and channel data-rates are computed in every time-slot following determination of available channels. Average channel fading over a long UAM trip duration can be considered constant. Therefore, CSI is assumed constant and independent~\cite{channelcsi}. Also, due to availability of a dominant LoS link, loss due to fading is lower than path-loss; so instantaneous channel variations can be ignored.}
%\textbf{add a footnote justifying the CSI can be performed and cite two papers and highlight that this paper studies the system for a long time period so the instantaneous channel variation (fading effect) are ignored and the focus is on the path loss effect of the channel.}

\par Each BS can transmit data radially outwards within a predefined envelope known as the `Fresnel' zone. Negligible transmission can happen skywards; beyond a threshold altitude, the path loss becomes proportional to the fourth power of distance. Cellular communication has both line-of-sight (LoS) and non-line-of-sight (NLoS) modes, each contributing additional average losses, denoted by $L_{LoS}$ and $L_{NLoS}$, respectively. For a given carrier frequency, $f_{cell}$, if the Euclidean distance between the eVTOL and the $i^{th}$ base-station is $di_{cell}^2$, the $i^{th}$ LoS and NLoS path losses $PL_i^{LoS}$ and $PL_i^{NLoS}$ are defined as follows:
\begin{align}\label{cell_PL}
    PL_i^{LoS} = 10\nu\log_{10}\Big(\frac{4\pi f_{cell} di_{cell}}{\sqrt{G_c}C}\Big) + L_{LoS}, ~~\text{[dB]}, \nonumber \\
    PL_i^{NLoS} = 10\nu\log_{10}\Big(\frac{4\pi f_{cell} di_{cell}}{\sqrt{G_c}C}\Big) + L_{NLoS}, ~~\text{[dB]},
\end{align}
where $\nu$ is the path-loss exponent, $G_C$ is the directive gain of the base-station antenna and $C$ is the speed of light. The eVTOL-BS distance of separation,  $di_{cell} = f(di_{horz}, di_{vert})$ is a function, $f$ of both planar distance parallel to earth's surface ($di_{horz}$) and the altitude of the eVTOL ($di_{vert}$) with respect to the $i^{th}$ cell. This region is defined by Fresnel's zone, which maintains a 3D-envelope around the BS where the cellular network strength is sufficient to sustain reliable communication.
The model given in \eqref{cell_PL}) computes the regular channel path loss when the UAM is located within the Fresnel's zone of the $i^{th}$ base-station. However, beyond the Fresnel's zone, as the UAM moves away from a base-station or gains altitude, the Euclidean distance, $di_{cell}$, becomes very large, and the path-loss increases rapidly~\cite{wirelessbook}.\\

It may be noted that all pose information are converted to grid coordinates for computation of the 3-D Euclidean distance, where a 3-D Euclidean distance is defined as a $2-norm$ of the vector difference of the locations transmitting and the receiving antennas. Combining the LoS and NLoS modes generates the effective path-loss, $PL_i^{cell}$, as shown in  \eqref{cell_PL_eff}. The probability factor, $\alpha$ is calculated according to \cite{power_alloc}:
\begin{equation}\label{cell_PL_eff}
 PL_i^{cell}= \alpha PL_i^{LoS} + (1-\alpha) PL_i^{NLoS}, ~~\text{[dB]}.
\end{equation}

 The effective channel gain is represented as $Gain_i^{cell} = \frac{1}{\sqrt{10^{PL_i^{cell}/10}}}$. Following Shannon's theorem, the cellular channel capacity or data-rate, $r_i^{cell}$, offered by the $i^{th}$ base-station can be computed as follows:
\begin{equation}\label{cell_data}
    r_i^{cell} = \omega_{cell}\log_{2}\Bigg(1+ \frac{PT_{cell}  \times |Gain_i^{cell}|^2}{\omega_{cell} \sigma_{cell}^2}\Bigg),
\end{equation}
where $\omega_{cell}$, $PT_{cell}$, and $\sigma_{cell}^2$ represent the cellular link bandwidth, transmission power, and average noise power respectively corresponding to the $i^{th}$ base-station. 
%\textbf{I dont think the omega\_cell in the denominator should be squared}

\subsubsection{UAM-Satellite Channel Model}
Data exchanged between eVTOL and LEO-satellite follows free-space propagation and hence suffers only free-space path-loss in LOS mode. The channel path loss between the eVTOL and the $j^{th}$ LEO-satellite is defined as $PL_j^{LEO}$.
\begin{equation}
    PL_j^{LEO} = 10n\log_{10}\Big(\frac{4\pi f_{LEO} dj_{LEO}}{\sqrt{G_S} C}\Big) + L_{LOS}, ~~\text{[dB]},
\end{equation}
where $f_{LEO}$ is the carrier frequency and $dj_{LEO}$ is the Euclidean distance between the $j^{th}$ LEO-satellite and the eVTOL at any sampling instant. The satellite antenna is assumed directive with gain $G_S$. The additional average loss suffered by the LoS link is $L_{LOS}$. Including fading characteristics, the effective channel gain is represented as $Gain_j^{LEO} = \frac{1}{\sqrt{10^{PL_j^{LEO}/10}}}$.The eVTOL to $j^{th}$ satellite data-rate is computed as follows:
\begin{equation}\label{satrate}
    r_j^{LEO} = \omega_{LEO}\log_{2}\Bigg(1+ \frac{PT_{LEO}  \times |Gain_j^{LEO}|^2}{\omega_{LEO} \sigma_{LEO}^2}\Bigg),
\end{equation}
where $\omega_{LEO}$, $PT_{LEO}$, and $\sigma_{LEO}^2$ denote the UAM-LEO satellite link bandwidth, transmission power, and average noise power, respectively

\section{In-Flight UAM Cellular-LEO Connectivity}
%\textbf{Add an introductory paragaph}
%Enhancing network coverage for UAM communication using cellular and satellite links can be established while minimizing data-loss due to the limited bandwidth during flights.
\subsection{Problem Formulation}
Assuming a fixed-size message $S(k)$ is transmitted at $k^{th}$ time-slot and the available data-rate at $k^{th}$ time-slot is $r(k)$, the cumulative data-loss ($CDL$) per trip can be represented as shown in \eqref{ps}. The data-loss is computed per time-slot and each time-slot being independent of another, the CDL can be computed at the end of the trip as follows:
\begin{equation}\label{ps} 
CDL= \min{\sum_{k=0}^{\check{K}} {\Big( S(k) -r(k) \delta t\Big)}}.
\end{equation}
%\textbf{the paragraph below should be in Section II}
At each time-slot, the data-rate $r(k)$ is a variable parameter that depends on the communication link allocated to the UAM.
We assume the cellular and the satellite channel designs are fixed, so the corresponding data rates are un-alterable. Therefore, data-loss can be reduced by minimizing `data-latency' (`latency' refers to the overall communication delay per UAM trip). We propose a link allocation scheme that evaluates the available communication channels in any time-slot and optimally assigns one link to the UAM. During take-off and landing, UAM altitude is within the cellular coverage area. Whereas, beyond the `Fresnel zone' the LEO-satellite network is available during stable flights. At the transition between the two networks, the UAM may be connected to either of the two networks, depending upon the altitude attained by the eVTOL.

In each time-slot, \textit{(a)} if the allocated link data rate is adequate to transmit the desired message, then the message is delivered without delay. If, \textit{(b)} the message $S(k)$ is completely transmitted at the $(k+1)^{th}$ time-sample, a successful data transmission with one unit delay occurs. However, \textit{(c)} if the desired message cannot be transmitted at the current data-rate on or before a pre-defined time expiry-limit, the communication protocol allows partial data transfer. Upon reaching the time expiry-limit, the remaining undelivered data is lost. In order to mathematically define the required transmission time for every message $S(k)$, we introduce the `delay' parameter as given in the following:
\begin{equation}\label{delay}
\Delta(k) := \frac{S(k)}{r(k)} \bigg| k \in \{1,2,...,\check{K}\},
\end{equation}
where the delay $\Delta(k)$ is upper-limited by $M \in \mathbb{I^+}$ time-slots, $M\delta t$ indicating expiry time-limit of undelivered data.

\subsection{Optimal Link Allocation}
Consider a generic situation where at time-sample $k$, the UAM is visible to $n \subset N_{cell}$ cellular BSs and $m \subset N_{sat}$ LEO satellites. Each cell offers a data-rate of $r_i, i \in \{1,2,...,n\}$ bps and each satellite provides a data-rate of $r_j, j \in \{1,2,...,m\}$ bps, respectively. Let us assume that each communication link incurs a Fixed Delay (FD) for data processing. Given so, the optimal link selection problem can be modeled as a Binary Integer Linear Program (BILP) formulated as in equation (\ref{BILP}), where, $\mathfrak{S}:=[0,1]$.
\begin{equation}\label{BILP}
\Delta^{opt}(k)= \min\limits_{X_i, Y_j \in \mathfrak{S}}{\sum_{i=1 \atop j=1}^{n \atop m} {\big(X_i \Delta^X_i + FD^X_i + Y_j \Delta^Y_j + FD^Y_j\big)}}.
\end{equation}
%\textbf{what is $F$ in (9)? It is not clear the use of the index i and j in (9), why under the $\min$ only the index i is used. }
Here, $\Delta^{opt}(k)$ is the optimal delay that will be incurred for delivering a message $S(k)$ transmitted in the $k^{th}$ time-slot. Superscripts $X$ and $Y$ distinguish the delays for cellular and satellite links, respectively, whereas, subscripts $i$ and $j$ are indicative of the corresponding base-station and satellite indices, visible in the $k^{th}$ time-slot. The fixed delays for both communication links have been assumed zero, so as to focus only on the link selection procedure. The coefficients $X_i$ and $Y_j$ are the optimization variables which will be assigned binary values in an optimal fashion for minimizing the delay $\Delta^{opt}(k)$. Since the coefficients $X_i$ and $Y_j$ have been formulated as weights of the communication links, the binary assignment must be constrained by the following equation (\ref{constraint}):
\begin{align}\label{constraint}
\sum_{i=1}^{n}{X_i} + \sum_{j=1}^{m}{Y_j} \leq 1, 
%\sum_{i=1}^{i=n}{X_i} + \sum_{j=1}^{j=m}{Y_j} > 0,
\end{align}
to ensure one and only one link is selected per time-slot %\textbf{it is not clear what is the purpose of the second line in (10), by definition, the expression is positive, the only case is all 0, which is possible in practice if you dont find any available link per time slot. I think we just need the first one to be  $\leq1$. It seems we want to force transmission because the objective function minimizing delay meaning simply I do not transmit anything so there is no delay}. 
The data-rate of the selected link (whose coefficient is 1) is the optimal link data-rate, $r^{opt}(k)$ in the $k^{th}$ time-slot.
\begin{figure}[h]
\centering
{\includegraphics[width=\columnwidth]{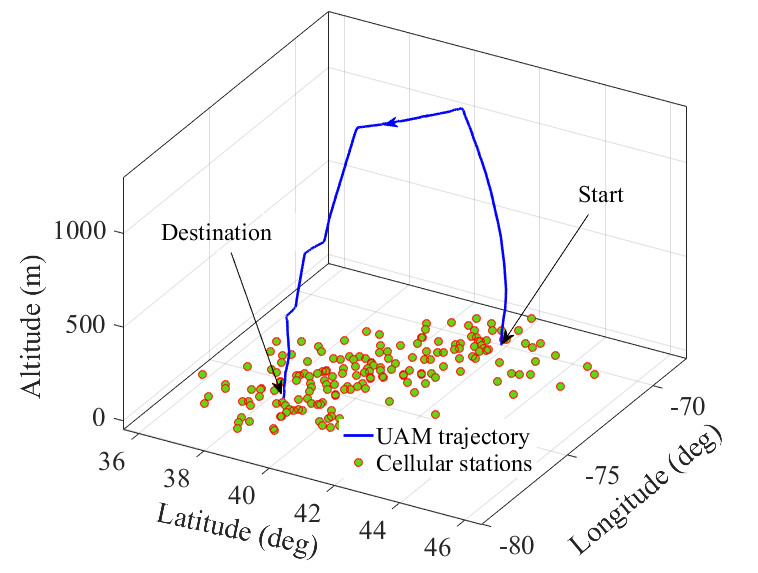}} 
\caption{eVTOL trajectory and the spatial distribution of cellular BSs. }
\label{Fig1}
\end{figure}
\noindent The cumulative delay per UAM trip can be expressed as follows: 
\begin{equation}\label{tripdelay}
  \text{Delay per trip}=\sum_{k=0}^{\check{K}} \Delta(k).
\end{equation}
Therefore, the optimal selection of a communication link in every time-slot results in the minimization of the total latency per trip, which in turn allocates the maximum available data-rate in every time-slot and hence minimizes the cumulative data-loss. In the case of partial data transfer, the delay need not be quantified. Instead, in each time-slot, the following Average Percentage Data-Loss (APDL) can be used as a performance metric, given, the time-base till $k^{th}$ time-sample is represented by parameter $p \in \{0,1,...,k\}$:
\begin{equation}\label{percentloss}
  \text{APDL} = \frac{\sum^k_0 (S(p) - r^{opt}(p)\Delta^{opt}(p)) \times 100}{\sum^k_0 S(p)}.
\end{equation}
%\textbf{it is not lcear why the sum is up to $k$? and the index inside the sum in $k$?}
The importance of the expiry time-limit parameter, $M \delta t$ becomes pronounced if the undelivered data has a provision to be stored and delivered later by a process termed as `caching'. For critical and time-sensitive messages, the accuracy and freshness of data may pose counter challenges, deliberating an appropriate numeric assignment to the parameter $M$. Caching has not been discussed in this paper and has left to a future extension of this work. Therefore, we assume $M=1$. 

\section{Results and Discussion}
The simulation scenario considered in this paper involves an eVTOL traveling from Boston to Washington D.C., covering an aerial distance of approximately 650 Km including take-off and landing. Due to vertical take-off and landing, time for taxi is negligible. Assuming an average speed of 163 Km/hr, the UAM trip duration is 4 hours, sampled every 5 seconds. The UAM trajectory is assumed to maintain a stable flight around 1300 Km above ground level for most of the trip duration. 184 cellular BSs have been placed along the UAM route such that at each time sample a cellular connectivity is available. Figure~\ref{Fig1} illustrates the distribution of BSs along the eVTOL route, which we consider a typical cellular network where BSs' antennas are tilted towards the ground.   
\begin{figure}[t]
\centering
{\includegraphics[width=\columnwidth]{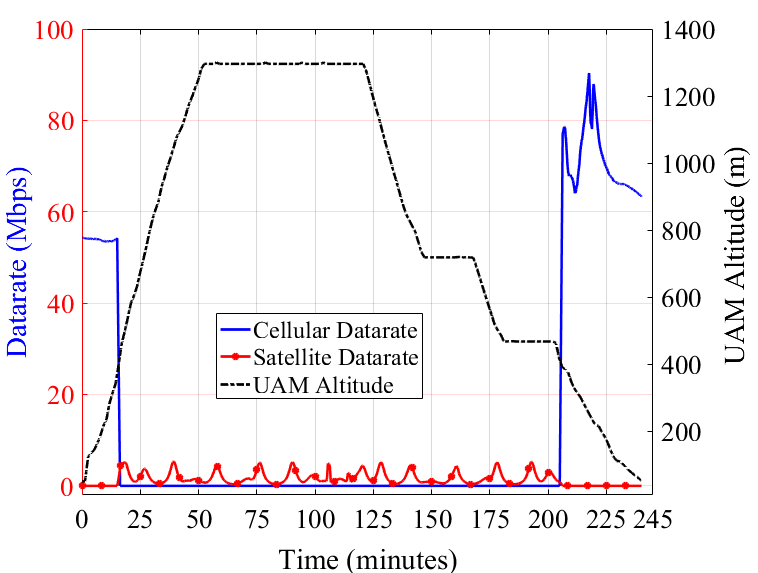}} 
\caption{UAM connectivity datarate under the hybrid cellular-LEO network. }
\label{Fig3}
\end{figure}
\setlength{\belowcaptionskip}{-12pt}
\begin{table}[t]
\centering
\label{Tab0} 
\caption{Satellite Orbital Parameters} % \hspace{-.8cm}
\addtolength{\tabcolsep}{-2pt}\begin{tabular}{|l r||l r|}
\hline
\textbf{Parameter} & \textbf{Value} & \textbf{Parameter} & \textbf{Value}\\ \hline
Semi-Major Axis (Km) & 7.23 $\times 10^3$ & Eccentricity (\%) & 0.0245  \\ \hline
Inclination (deg)& 53.94 & Period (hours) & 1.7\\ \hline
Argument of Periapsis (deg) & 169.98 & True Anomaly (deg) & 306.66 \\ \hline
\end{tabular}
\end{table}
\setlength{\belowcaptionskip}{0pt}
However, cellular coverage is assumed to effective within a radius of 2 km and an altitude of 400 m above the BS antenna. Beyond 400 m the UAM communicates with the LEO-satellite network. Below, Figure \ref{Fig3}
shows the cellular and satellite data-rates in relation to the UAM altitude profile. 

\begin{figure}[t]
\centering
{\includegraphics[width=8cm]{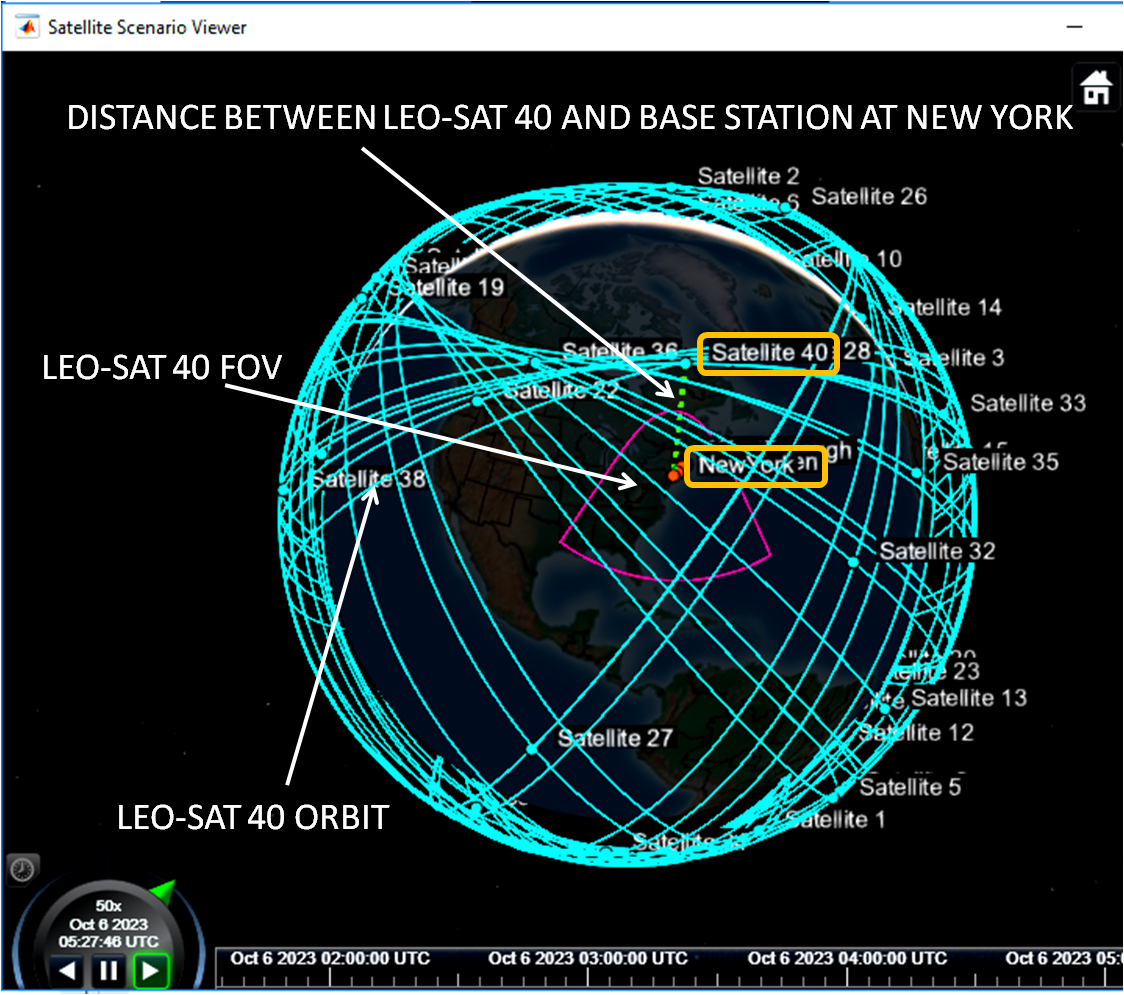}} 
\caption{Satellite Scenario View - BS `NEW YORK' observed from a visible satellite (ID 40) and calculation of distance between them while orbiting the Earth.}
\label{Fig4}
\end{figure}

The LEO satellite constellation file is available as a `two-line element' dataset in Matlab2023b. Satellite Communication Toolbox offers a satellite scenario simulator which allows the user to model satellites and cellular BSs and study interactions between them. This tool enables the user to define the satellite orbit parameters such axis, eccentricity, inclination, propagator model, as well as simulation date-time objects. An exemplary set of parameters for satellite ID 40 is illustrated in Table I. A constellation of 40 LEO satellites, whose position and velocity are generated in every time slot of the UAM trip duration, has been considered for simulation. The UAM data have been generated from the flight statistics provided by \textit{Flightaware}, with changes made in the altitude and velocity for UAM. An instance of the satellite scenario generator showing visibility of a base-station from a specific satellite and calculation of the distance between them has been depicted in Figure~\ref{Fig4}.
%\vspace{0.5cm}

LEO carrier frequency and bandwidth have been modeled as Starlink satellites mentioned in \cite{starlink}. The cellular BSs represent a standard 5G network. Since the LEO-UAM connectivity follows free space propagation, the satellite channel operates only along the line-of-sight, incurring only LoS losses. Whereas, UAM-cellular link suffers from both LoS and NLoS losses. Both satellite and cellular antennas are considered directive, the satellite antenna having higher gain and lower transmission power due to input power constraints. We assume both communication links utilize Frequency Division Multiple Access (FDMA) methods to increase the corresponding channel bandwidth by twelve-fold. Table II gives a comprehensive overview of the simulation parameters used in this paper.
\begin{table}[t]
\centering
\label{Tab1} 
\caption{Simulation parameters} % \hspace{-.8cm}
\addtolength{\tabcolsep}{-5pt}\begin{tabular}{|l|c| r|}
\hline
\textbf{Parameter} & \textbf{Sat-Comm} & \textbf{Cellular}\\ \hline
Carrier frequency, $f$ (MHz) & 11000 & 850  \\ 
Bandwidth, $\omega$ (KHz) & 400 & 200 \\ 
FDMA Channels & 12 & 12 \\ 
LoS path loss, $L_{\text{LoS}}$ (dB) & 0.1 & 1  \\ 
NLoS path loss, $L_{\text{NLoS}}$ (dB) & \text{Nil} & 20 \\ 
Transmitter power (W) & 8  & 10\\ 
Antenna Gain (dB) & 20 & 12\\ 
Noise power, (dBm/Hz) & -174 & -174 \\ \hline
\end{tabular}
\end{table}

\begin{figure}[t]
\centering
{\includegraphics[width=8.5cm]{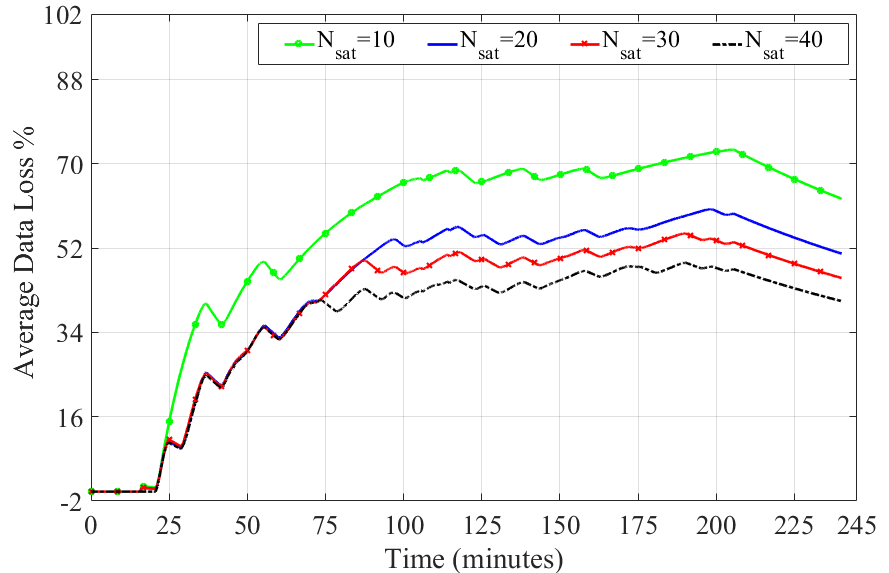}} 
\caption{Average data-loss percentage at every time-slot decreases with increase in constellation size.}
\label{Fig5}
\end{figure}

For a given constellation size $m$, the average data-loss percentage has been observed to increase until stable flight conditions are attained. Between 100-200 minutes of Figure \ref{Fig5}, it has been shown that the data-loss percentage becomes roughly constant, indicating consistent satellite visibility. However, the satellite data-rate is much lower compared to the cellular data-rate. Because of this, during take-off (0-25 minutes) and landing (210-245 minutes), along with the transitioning buffer time,the data-loss is either zero or decreasing. With an increase in the constellation size, the average data-loss percentage falls significantly from 75\% to 40\%.
\setlength{\belowcaptionskip}{1pt}
\setlength{\abovecaptionskip}{2pt}
\begin{figure}[t]
        \centering
        \begin{subfigure}[t]{0.5\textwidth}
                \centering
                \includegraphics[width=8.5cm]{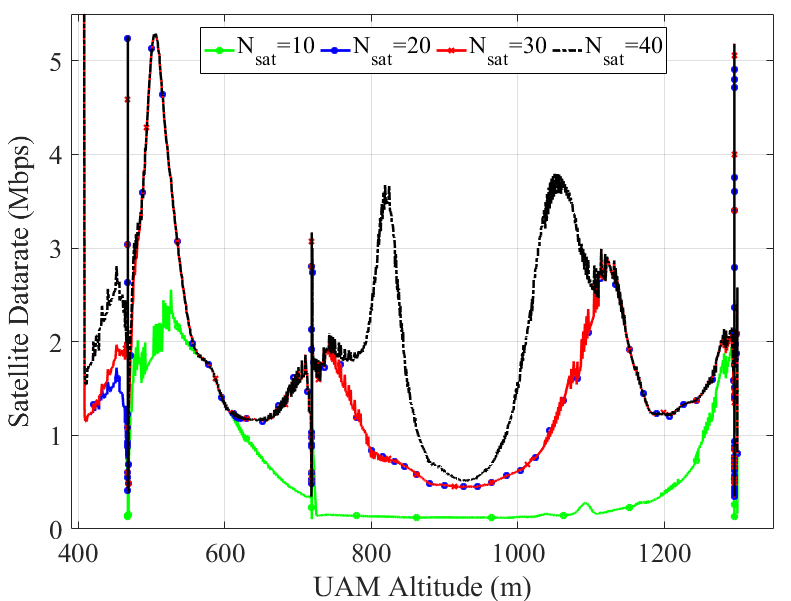}
                \caption{UAM-LEO satellite data-rate increases with more number of satellites during stable flight.}
                \label{Fig6}
        \end{subfigure}
        \begin{subfigure}[t]{0.5\textwidth}
                \centering
                \includegraphics[width=8.5cm]{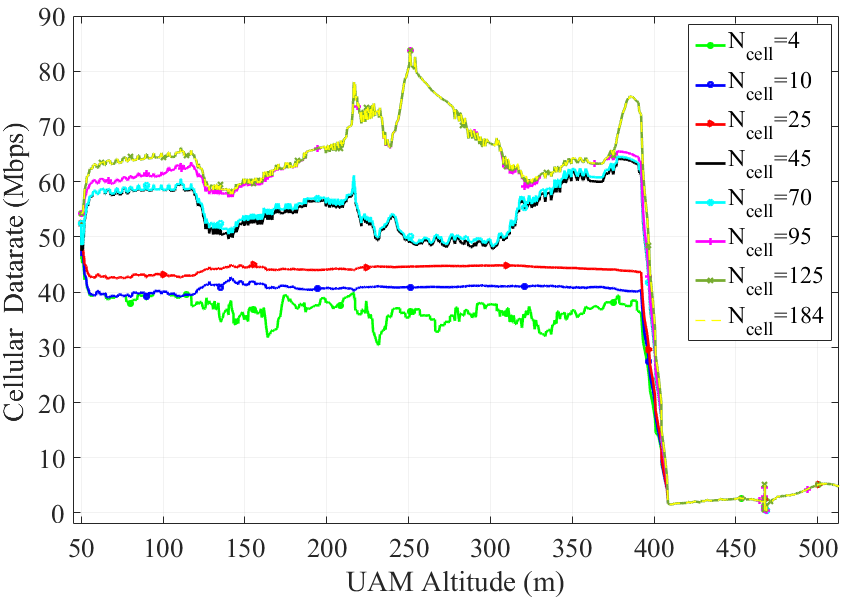}
                \caption{UAM-cellular data-rate increases with more number of BSs, only during take-off and landing. Beyond the 400m mark the Fresnel's zone ends and channel data-rate belongs to allocated satellite link. }
                \label{Fig7}
        \end{subfigure}
        \caption{Variations in satellite and cellular data-rate per trip.}\label{datarate_var}
\end{figure}
As discussed earlier, since minimization of data-loss has a direct bearing with the assignment of the link with maximum available data-rate, it is worth investigating how the satellite and the cellular data-rates change with the corresponding network size. The results shared in Figures~\ref{Fig6} and \ref{Fig7} can also be utilized to determine the optimal number of cellular stations and satellites to be considered for effective connectivity. Figure~\ref{Fig6} illustrates that a significant increase in data-rate can be seen if $m=30$ satellites are considered instead of $m=10$, but much difference in performance is not observed if the constellation size is increased further to $m= 40$. Satellite communication being expensive an optimal network size may just be taken as 30. Similarly, variation with number of BSs results in a rapid increase in data-rate from $n=25$ to $n=125$, but not much if further increased to $n=184$. On the other hand, if the desired message is transmittable without delay with 40 Mbps, we can limit the cellular network size to $n=25$ BSs.

%\vspace{1cm}
\section{Conclusion}
In this paper, we have investigated a strategy for ensuring connectivity for UAM vehicles via either cellular or satellite networks by minimizing UAM data-loss. This has been achieved by optimal allocation of cellular and LEO satellite communication links at each time-sample using binary integer linear programming. The proposed algorithm can be extended to serve complex data communication involving varying data-priority, and time-sensitivity using the concept of data-caching.
\label{conclusion}

\balance
\nocite{*}
\bibliographystyle{IEEEtran}
\bibliography{uam.bib}

\vfill
\end{document}